Classification: Biological sciences; Plant Biology
Title: Comprehensive analysis of imprinted genes in maize reveals limited conservation with other species and allelic variation for imprinting.
Running title: Comprehensive analysis of imprinting in maize

Authors
Amanda J. Waters[1], Paul Bilinski[2], Steve R. Eichten[1], Matthew W. Vaughn[3], Jeffrey Ross-Ibarra[2,4], Mary Gehring[5], Nathan M. Springer[1§]

[1] Microbial and Plant Genomics Institute; Department of Plant Biology University of Minnesota, Saint Paul MN 55108
[2] Department of Plant Sciences, University of California, Davis, CA 95616
[3] Texas Advanced Computing Center, University of Texas-Austin; Austin TX 78758
[4] The Genome Center and Center for Population Biology, University of California, Davis, CA. 95616
[5] Whitehead Institute for Biomedical Research, Cambridge, MA 02142; Department of Biology, Massachusetts Institute of Technology, Cambridge, MA 02139

[§]Author for correspondence

Corresponding author:
Nathan Springer
Microbial and Plant Genomics Institute
Department of Plant Biology
University of Minnesota
250 Biosciences Center
1445 Gortner Ave
Saint Paul MN 55108
Phone: (612)624-6241
springer@umn.edu

**Abstract**:

In plants, a subset of genes exhibit imprinting in endosperm tissue such that expression is primarily from the maternal or paternal allele. Imprinting may arise as a consequence of mechanisms for silencing of transposons during reproduction, and in some cases imprinted expression of particular genes may provide a selective advantage such that it is conserved across species. Separate mechanisms for the origin of imprinted expression patterns and maintenance of these patterns may result in substantial variation in the targets of imprinting in different species. Here we present deep sequencing of RNAs isolated from reciprocal crosses of four diverse maize genotypes, providing a comprehensive analysis of imprinting in maize that allows evaluation of imprinting at more than 95% of endosperm-expressed genes. We find that over 500 genes exhibit statistically significant parent-of-origin effects in maize endosperm tissue, but focused our analyses on a subset of these genes that had >90% expression from the maternal allele (69 genes) or from the paternal allele (108 genes) in at least one reciprocal cross. Over 10% of imprinted genes show evidence of allelic variation for imprinting. A comparison of imprinting in maize and rice reveals that only 13% of genes with syntenic orthologs in both species exhibit conserved imprinting. Genes that exhibit conserved imprinting in maize relative to rice have elevated dN/dS ratios compared to other imprinted genes, suggesting a history of more rapid evolution. Together, these data suggest that imprinting only has functional relevance at a subset of loci that currently exhibit imprinting in maize.

**Introduction**:

Imprinting describes a biased expression of alleles that depends upon the parent of origin. Imprinting is observed in both flowering plants and mammals (1-3). Most mammalian imprinted genes occur in clusters with other imprinted genes and imprinting is often conserved at well-characterized imprinted genes among mammals (1, 4). In plants, imprinted genes exhibit relatively little clustering and imprinting is largely confined to the endosperm, a triploid tissue that contains two maternal genomes and a single paternal genome. The endosperm provides an energy source for germinating seeds and, as the majority of harvested grain consists of endosperm tissue, a major source of calories in the human diet. A better understanding of imprinting will shed further light on the mechanisms of epigenetic gene regulation and endosperm development and could provide an avenue for altering reproductive processes or seed quality in plants.

Despite a widespread interest in imprinting and its potential importance, the function of most imprinted genes is not well-characterized in plants and imprinting has only recently been assayed on a genome-wide level. Imprinting is reflected in parentally biased allele-specific expression in the endosperm tissue of intraspecific reciprocal hybrids. A quantitative method for detecting the relative expression of two alleles that have nearly identical sequences is required to find such an effect, traditionally limiting analysis to a handful of imprinted genes identified based on phenotype or through targeted analyses (5-9). The implementation of deep sequencing of RNA molecules (RNAseq) has allowed detection of additional imprinted genes (10-15). In each of these studies, allele-specific expression levels were monitored for a single cross of two parents in *Arabidopsis*, maize or rice. This allowed for the analysis of imprinting in 50-58% of genes expressed in endosperm tissue. In each species there is evidence for several hundred imprinted genes with similar numbers of maternally expressed genes (MEGs) and paternally expressed genes (PEGs), but comparisons among flowering plants (12-13, 16-17) have revealed limited overlap in the genes that are imprinted among species.

There has been considerable speculation on the mechanisms that might lead to the origin of imprinted expression for a particular allele as well as the evolutionary mechanisms that would lead to the maintenance of imprinted expression (16, 18-20). Recent studies have been interpreted to suggest that imprinting may arise due to programmed release of heterochromatic silencing marks in specific nuclei of the male and female gametophytes (21). Plant gametophytes are multi-nucleate structures. The male gametophyte includes a vegetative nucleus and two sperm nuclei. The female gametophyte often has multiple cells including the haploid egg cell (which is fertilized by a sperm nuclei to generate the embryo) and the diploid central cell (which is fertilized by a sperm cell to generate the endosperm) (22). The loss of DNA methylation before fertilization leads to an epigenetic asymmetry in the endosperm because the maternal genomes (from the central cell) have been demethylated while the paternal genome (from a sperm nucleus) retains normal levels of methylation. Programmed DNA demethylation might result from the generation of siRNAs that could reinforce transposon silencing in adjacent cell types (egg and sperm cells) that contribute genetic material to the next generation (23). It has been hypothesized that this process, while targeted to transposons, could inadvertently influence nearby genes, resulting in imprinted expression (19). In support of this idea, several well-characterized imprinted genes contain transposon sequences in adjacent regions (5, 24-25). The potential for transposons to contribute to the origin of imprinted expression for nearby genes may result in examples of imprinting that do not provide a selective advantage and would not be expected to persist over evolutionary time. Because imprinting at such loci would be of limited functional relevance and dependent on the presence of a transposable element, we might also expect to observe substantial allelic variation for imprinting within a species. Indeed, several of the first characterized examples of imprinting in maize exhibit allelic variation such that certain alleles are imprinted while others are not (26-27).

While imprinting of a gene may arise inadvertently due to the regulation of nearby transposons, it is also possible that parent-of-origin specific expression could in some instances provide a selective advantage. The kinship theory (18) suggests that maternally expressed genes would restrict growth or limit the flow of resources to offspring while paternally expressed genes might function to promote offspring growth. There are examples of imprinted genes that appear to exhibit these functions (28), but there is no clear evidence for these predicted functions in the annotations of the full set of previously identified MEGs or PEGs (3). Genes that are subject to parental conflict might be expected to exhibit signatures of positive selection (20, 29). For some imprinted genes, such as the *Arabidopsis* seed size locus MEDEA, potential evidence of positive selection has been found in some cases (30-31), but not others (32).

The presence of imprinting for a particular gene is often assumed to have functional relevance. While this may be the case for a subset of genes, the potential for inadvertent acquisition of imprinting as a result of nearby transposon influences could result in numerous examples of imprinting that have limited functional relevance and thus show intra- or inter-specific variation in imprinting. To distinguish between these possibilities and evaluate the functional importance of imprinting, we analyzed imprinting in multiple diverse genotypes of maize. Reciprocal crosses among four genotypes provided the ability to survey imprinting at over 95% of the genes expressed in endosperm tissue. We documented numerous differences in the regulation and patterns of maternally expressed genes (MEGs) and paternally expressed genes (PEGs) and find that only a subset of imprinted genes show conserved imprinting in maize and rice and that these genes show evidence of distinct selective pressures. Comparison of imprinting in different haplotypes within maize reveals allelic variation for imprinting, further suggesting that imprinting may have limited functional consequence for many maize genes.

**Results**:

Deep sequencing of RNA isolated from 14 day after pollination (DAP) endosperm tissue of five reciprocal hybrid pairs was performed to identify imprinted genes. This intermediate stage of endosperm development was selected because it is before major starch accumulation but after endosperm cellularization. Analysis of this stage also reduces the effects of transient imprinting that has been observed for some genes at earlier stages of development (33) as well as the contribution of transcripts from the gametes. The five reciprocal hybrids included one previously analyzed dataset for the cross of inbred lines B73 x Mo17 (13) as well as four new reciprocal hybrids generated by crossing inbred lines Ki11 and Oh43 with both B73 and Mo17 (Table 1). These additional genotypes were selected because whole-genome resequencing provided detailed SNP calls (34) and because they represent diverse genotypes (35).

A large number of reads (180-210 million) were recovered for each of the 10 genotypes and were analyzed to study gene and allelic expression patterns (see Methods, Fig. S1 for details). The number of reads that mapped to each allele was summed across all SNPs for a transcript. Only transcripts that had at least 10 reads that could be assigned to a particular allele in each direction of the reciprocal cross were analyzed, resulting in allelic expression data for between 5,851 and 13,478 genes in each cross (Table 1: Fig. S1). In total 18,284 genes (95% of genes expressed in 14 DAP endosperm) had allele-specific expression data in at least one of the five reciprocal hybrid pairs (Table 1; Fig.s 1A and S2). In maternally expressed genes (MEGs), the maternal allele will be preferentially expressed, revealing a higher than expected proportion of the maternal allele in both directions of the cross. Paternally expressed genes (PEGs), in contrast, will exhibit low levels of the maternal allele in both directions of the cross. Genes that exhibit consistent bias for the allele from one genotype, independent of parent of origin, reflect cis-regulatory allelic variation.

*Comprehensive discovery of maize imprinted genes*

A combination of statistical significance and proportion filters were implemented to identify and classify differing levels of MEGs and PEGs (Fig. S1). We assigned different levels of imprinting to parentally biased genes to compare imprinting strength between species and between different types of genes in a more nuanced manner. Moderate MEGs / PEGs were defined as having significant allelic bias ($\chi^2$<0.05) and >80% of transcripts from the maternal parent (MEGs) or >60% of the transcripts from the paternal parent (PEGs) (red shaded areas in Fig. 1A) in both directions of a reciprocal cross. These criteria are slightly different for MEGs and PEGs because the expected value in triploid endosperm is 2:1 instead of 1:1. Strong MEGs and PEGs were defined as having significant allelic bias ($\chi^2$<0.01) and >90% of transcripts from the maternal parent (MEGs) or paternal parent (PEGs) (blue area in Fig. 1A). Complete MEGs or PEGs have >99% of the transcripts derived from the maternal or paternal allele, respectively. In addition, we also identified a series of genes for which there is strong allelic bias (at least 95% reads from one parent) in one direction of the cross but not in the reciprocal hybrid as potentially indicative of allelic variation for imprinting (green shaded areas in Fig. 1A).

The number of genes classified as moderate, strong or complete MEGs and PEGs varied for each genotype (Table 1, Fig. 1B) in large part due to differences in the number of genes with polymorphisms. For the majority of subsequent analyses, only the genes that were classified as strong or complete MEGs/PEGs were used. There are a total of 108 non-redundant strong PEGs, including 31 (28%) examples that were classified as complete PEGs in at least one genotype (Table 1, Data S1). Additional filtering criteria were applied to MEGs to remove genes that might exhibit maternal bias due to contamination of maternally derived tissues. RNA-seq data from a B73 expression atlas (36) was used to identify MEGs that may be the result of maternal contamination, resulting in a filtered list of 69 non-redundant strong MEGs, with a larger number (37; 54%) showing complete imprinting than seen in PEGs (Table 1; Data S2).

Quantitative SNP assays designed using the Sequenom MassArray platform (37) were used to validate imprinting for 13 MEGs and 13 PEGs (Table S1). These assays are based on a single SNP for each gene and could only be used to assess imprinting in the crosses that were polymorphic for the targeted SNP. The analysis of allele-specific expression in a different 14 DAP endosperm sample for the same set of five reciprocal crosses confirmed imprinting in the majority of samples for both MEGs (23/24) and PEGs (28/28). The one allele that was not validated showed imprinted expression in one direction of the cross but bi-allelic expression in the reciprocal hybrid. The same quantitative SNP assays were also used to assess whether imprinting for these genes was also detected in several other genotypes (NC358, Ms71, and M162W) that were reciprocally crossed with B73 and Mo17. Most of these genes were imprinted in each of the other genotypes that were tested, with the exception of one locus (GRMZM2G020302) that was imprinted in both M162W and Ms71 but did not show imprinting in NC358 (Table S1). Finally, the quantitative SNP assays were also used to assess whether imprinted expression was maintained at earlier and later stages of endosperm development. Imprinting was consistently observed for 26/26 MEGs and 25/26 PEGs at 12 DAP, 14 DAP, 16 DAP, and 20 DAP samples of B73xMo17, B73xNC358 and Mo17xNC358 (Table S1).

*Characterization of maize imprinted genes*

Several plant imprinted genes have expression that is restricted to the endosperm. This endosperm-specific expression could be because these genes have specific functions in the endosperm, or because it is beneficial to silence these genes in somatic tissues. Endosperm-specific expression of a single allele would then simply be a consequence of epigenome reprogramming in the gametophytes (20). While it has been suggested that endosperm-specific expression is a general feature of imprinted genes we find that only a subset of MEGs and PEGs exhibit preferential expression in endosperm relative to other tissues in maize (Fig. 2). These include genes that are only detectable in endosperm as well as genes that are significantly higher expressed (>5-fold difference) in endosperm compared to other plant tissues. The majority of the MEGs (68%) are preferentially expressed in endosperm while only 26% of the PEGs are preferentially expressed in endosperm (Fig. 2). Many of the MEGs exhibit increasing levels of expression during endosperm development, suggesting that these genes are actively transcribed in endosperm tissue as opposed to being stable, maternally inherited transcripts. MEGs and PEGs exhibit a range of expression levels in 14 DAP endosperm tissue (Fig. S3A). The genes with preferential expression in endosperm tend to have higher expression levels in endosperm than imprinted genes that are also expressed in vegetative tissues (Fig. S3A).

Consistent with previous work (14, 38), we find that PEGs are more likely to be targets for histone methylation than MEGs. Only 5 of 69 of the MEGs exhibit H3K27me3 in endosperm tissue (Data S2), in contrast to 87 of the 108 PEGs (Data S1). The 87 PEGs that are marked with H3K27me3 in endosperm tissue include 64 genes with expression in vegetative tissues and 23 genes with preferential expression in endosperm (Data S1). Only 8% of the 64 PEGs that are expressed in vegetative tissues exhibit H3K27me3 in the four vegetative tissues analyzed, while 65% of the 23 PEGs with preferential expression in endosperm are marked by H3K27me3 in at least three of the four vegetative tissues that were analyzed (Data S1).

MEGs and PEGs also differ in their conservation between species and their annotation. The frequency of PEGs with syntenic orthologs in rice (39) was much higher (83%) than MEGs (46%) (Data S1-2). Similarly, the proportion of PEGs with high sequence similarity (E<1E-50) to an Arabidopsis gene (61%) was higher than the proportion of MEGs (36%) (Data S1-2). Overrepresentation of functional categories of GO annotations were investigated for the 62 PEGs and 23 MEGs that had high sequence similarity to Arabidopsis (E-score <10E-50) using BinGO (40). PEGs exhibit significant (p<0.05) enrichment for GO terms including developmental process, response to stimulus and macromolecule modification. MEGs exhibit significant (p<0.05) enrichment for macromolecule modification, cellular metabolic process, and kinase activity.

*Allelic variation for imprinting*

Several of the earliest examples of imprinted loci exhibited imprinting for some alleles but not for others (26, 27). Our analysis of multiple maize genotypes provides an opportunity to comprehensively assess allelic variation in imprinting (Fig.3A). In general, when data were available for multiple crosses, many (88%) genes that exhibit imprinting in one cross were also imprinted in the other crosses, but there are examples in which genes imprinted in one cross display allelic variation for imprinting in another cross (Fig. 3, S4). We identified 17 genes (8 PEGs and 9 MEGs) that showed consistent patterns of allelic variation in imprinting (Fig. 3B, Data S3). In each case, the same allele exhibited a lack-of-imprinting in multiple crosses or was confirmed by quantitative SNP assays. Gene GRMZM2G384780, for example, shows complete maternal imprinting in the Mo17/Oh43 cross, but crosses involving B73 fail to silence the B73 allele when paternally inherited (Fig. 3C). Similar variation can be observed for other MEGs (Fig. S4) and PEGs (Fig. 3D and S4). The PEG (GRMZM2G106222) shows expression of the maternal allele only when Oh43 is the maternal parent (Fig. 3D). A quantitative SNP assay was used to confirm the allele-specific imprinting for GRMZM2G106222 (Fig. 3D) and one of the other genes (GRMZM2G020302) exhibits allelic variation for imprinting in NC358 (Table S1). Overall, these data provide evidence for standing allelic variation for imprinting at 12% (17/144, the total is the number of genes with data in at least two sets of reciprocal crosses) of the imprinted genes even though we only assayed up to four haplotypes for each locus.

*Conservation of imprinting between species*

If imprinting plays a similar functional role in all flowering plant species, regardless of differences in endosperm growth or development, then it might be expected that there would be strong conservation for the targets of imprinting. Previous work has found only 5-10 examples of conserved imprinting between species (3, 12-13), but has had limited comparative power due to the use of only a single cross in which not all genes may show polymorphism. The availability of a comprehensive list of MEGs and PEGs analyzed in multiple crosses in maize and information on syntenic gene relationships in rice provided an opportunity to investigate the conservation of imprinting in monocots in more detail. There are 58 maize PEGs and 27 maize MEGs that have syntenic orthologs in rice that were assessed for imprinting by Luo et al. (2011). Of these, 9 PEGs and 3 MEGs show imprinting for both the maize and rice syntenic orthologs (Fig. S5C) and an additional 2 PEGs and 1 MEG that have imprinting for a closely related rice gene not located at a syntenic genomic position (Fig. S5A). This is a relatively low level of conservation but is significantly higher than expected by chance ($\chi^2$, p<0.001). There are also 3 moderate MEGs and 8 moderate PEGs that show imprinting in their corresponding syntenic rice gene (Fig. S5D), and a low, but statistically significant ($\chi^2$, p<0.001) level of conservation for imprinting of related sequences (not necessarily syntenic) in maize and *Arabidopsis* (Fig. S5B). Genes with conserved imprinting in maize and rice include a variety of annotations (Fig. S5C-D). Two of these, encoding an ARID/BRIGHT DNA binding domain protein and a flavin-binding monooxygenase protein, also show imprinting for related sequences in *Arabidopsis*.

Finally, we analyzed the conservation of imprinting between paralogs from the recent whole-genome duplication event in maize. Following an allopolyploid whole-genome duplication event 5-12 million years ago (41) subsequent rearrangements and fractionation have resulted in varying patterns of retention and loss (39, 42) of syntenic paralogs. A larger proportion of PEGs (73 genes, 68%) than MEGs (31 genes, 45 %) are found in one of the two syntenic blocks assigned to subgenomes (Fisher's exact test two-tailed p-value = 0.005; Data S1 and S2). The larger number of MEGs outside of syntenic blocks may be due to recent duplication: 17% (12/69) of MEGs show greater than 95% homology via BLAST to another gene in the genome compared to only 6% (6/108 of PEGs

(Fisher's exact test two-tailed p = 0.0037) (Table S3). For those MEGs and PEGs found in either subgenome, both groups show similar ratios of genes with retained syntenic duplicates (7/31 MEGS and 18/73 PEGS, p = 1.0) (Table S3). Of the 7 MEGs with retained duplicates in both subgenomes, two of the duplicates exhibit moderate imprinting, two are not imprinted but are expressed in the endosperm, and three are not expressed in the endosperm (Data S2). Among the 18 PEGs with retained duplicates, 10 are imprinted, 7 are expressed in the endosperm but not imprinted, and 1 is not expressed in the endosperm (Data S1).

*Evolutionary genetics of imprinting*

To further investigate the evolution of imprinted loci, we took advantage of recent whole-genome analyses of maize and teosinte (43) to compare patterns of genetic diversity in imprinted and non-imprinted genes. In spite of the likelihood of selection on kernel traits (including endosperm) during recent maize evolution, we find no evidence that imprinted loci are enriched in regions targeted by selection during domestication or subsequent improvement. Moreover, imprinted genes themselves show few signs of selection, with values of nucleotide and haplotype diversity generally similar to genome-wide trends (Table S2). The only exception to this trend can be found in the paucity of high-frequency derived mutations seen in MEGs (median normalized Fay and Wu H = 1.11, Wilcoxon rank sum test p-value=0.0028), perhaps suggesting weak purifying selection.

We further evaluated the evolutionary importance of imprinted genes by comparing the ratio of non-synonymous to synonymous substitutions (dN/dS) between maize, rice, and sorghum (Fig. 4). Genes with conserved imprinting show higher dN/dS values than both non-conserved imprinted genes (Wilcoxon rank, p<0.01) and all genes tested (Wilcoxon rank, p<0.01), though non-conserved imprinted genes differ from other tested genes in maize-rice and maize-sorghum comparisons (Wilcoxon rank, p<0.01) (Fig. 4). Codon-based analysis of dN/dS in both conserved and non-conserved imprinted loci revealed the predominant effects of purifying selection across both classes of loci (Fig. S6). However, approximately half of conserved imprinted genes showed evidence of positive selection on at least one codon, compared to only one non-conserved imprinted gene that showed any evidence for positive selection (Fig. S6A).

**Discussion:**

Our analysis of allele-specific expression in multiple crosses of maize is the most comprehensive study of imprinting in any plant species to date. Over 95% of the genes that are expressed in endosperm could be tested for imprinting due to the presence of polymorphisms in at least one of the crosses. There is evidence for several hundred genes that show consistent parent-of-origin effects in at least one of the crosses, and a substantial subset of these moderate MEGs or PEGs exhibit strong or complete imprinting. The availability of a relatively complete set of imprinted genes for maize provided an opportunity to examine the level of conservation of imprinting both within and between species.

Imprinted genes are often treated as a single class in the literature. However, there are a number of differences between MEGs and PEGs that suggest that these different types of imprinting might reflect different processes and play different roles. MEGs are much more likely to exhibit endosperm-specific expression than PEGs, more likely to lack predicted function, and much less frequently associated with H3K27me3. There are also differences in the conservation of MEGs and PEGs between species. The maize PEGs have fewer recent duplications, and are more likely to have a retained a syntenic ortholog in rice and highly similar sequence in Arabidopsis. In addition, there are more examples of conserved imprinting in maize and rice, or between maize paralogs, for the PEGs.

The initial discovery of imprinting was based on studies of the R locus in maize (6), which exhibits allelic variation for imprinting (26). There are several other examples of potential allelic variation for imprinting in maize (33), but there have been few studies that assess imprinting for multiple alleles within a species. Our data reveal that over 10% of the genes with strong imprinting show allelic variation among the four maize haplotypes surveyed. This rate would undoubtedly increase if additional haplotypes were tested: we found at least one example of a gene for which the four alleles used for RNAseq were all imprinted but at least one additional genotype tested by a gene-specific assay was not (Table S1). This allelic variation in imprinting may reflect differences in transposon content near maize genes. Studies of haplotype structure variation in maize (44) provide evidence for substantial allelic variation in the type of repetitive elements surrounding genes. The lack of allelic variation for imprinting of genes showing conserved imprinting in rice also suggests that such variation may not be functional, but instead might simply reflect the inadvertent influence of polymorphic transposons upon nearby genes. Further study of the specific haplotypes present at alleles that are imprinted or not imprinted may shed further light on the genetic changes that contribute to imprinted expression.

We investigated the conservation of imprinting between two monocots with persistent endosperm, maize and rice. In total we identified 88 imprinted maize genes with a syntenic rice gene evaluated by Luo et al (2011), but only 12 exhibit conserved imprinting. While higher than expected by chance alone, this limited number suggests that conservation of imprinting over longer periods of evolutionary time is not common. It is important to note, however, that there are likely more than 12 examples of conserved imprinting in maize and rice because a number of loci, such as the rice ortholog of the maize MEG *Mez1* (32), could not be tested in rice due to a lack of polymorphisms (12). Nonetheless, genes with conserved imprinting tend to show elevated dN/dS ratios. The elevated dN/dS ratios could be the result of weak purifying selection or positive selection for a portion of the gene. The finding that a number of the conserved imprinted genes show evidence of positive selection for at least one codon is consistent with an important functional role or perhaps even their involvement in genomic conflict. The limited conservation of imprinting among species also may help to guide future functional studies. We might hypothesize the genes with conserved imprinting among species play important functional roles in regulating seed development and growth and would be useful targets for reverse-genetic analysis. In contrast, the genes with imprinting only in certain species may reflect unique reproductive strategies in those species or could result from inadvertent imprinting due to transposon variation which would not result in functionally relevant imprinting.

**Methods**:

*RNA-seq analysis*: Two ears of reciprocal F1 hybrid crosses of B73xMo17, B73xKi11, Mo17xKi11, B73xOh43, and Mo17xOh43 were collected. RNA isolated from 14 DAP endosperm tissue was sequenced using the Illumina HiSeq-2500 platform. Reads were aligned to the 39,540 genes in the filtered gene set using Tophat aligner (45), from which FPKM- fragments per kilobase per million reads and allele specific expression rates were calculated. RNA-*seq* reads are deposited at the NCBI SRA under accession (in progress) (see supplemental methods for details.)

*Allelic variation detection*: Allele specific read counts or Sequenom data for each set of reciprocal crosses were analyzed to discover genes that exhibit allelic variation of imprinting. Genes with at least 20 RNA-seq reads were run through a pipeline that pulls the maize gene ID for genes that showed allelic variation of imprinting in at least two sets of reciprocal crosses, and identifies which alleles are not imprinted at the locus. Additionally, data from the Sequenom assay validated a subset of genes that exhibit allelic variation of imprinting discovered in Waters et al (2011).

*Quantitative SNP assays*: Quantitative SNP assays (Sequenom MassArray) were used to validate imprinted genes and assess imprinting across additional genotypes or over a time course of seed development (see supplemental Methods for details).

*Annotation and comparative genomics of imprinted genes:* Maize syntenic orthologs in rice and retained whole-genome duplicates were identified based on the criteria outlined and database created from Schnable et al (2012). Maize imprinted genes with high sequence similarity (E-score >10E-50) to *Arabidopsis* were used to assess GO enrichment for functional categories using BinGO (40). GO enrichment categories were identified by being significant (p<0.05) and having at least 5 genes in each category.

*Diversity and divergence analyses:* Population genetic data from Hufford et al (2012) for a total of 14,982 (all genes with allelic expression data in endosperm tissue for at least one reciprocal cross) genes were included in our analysis, including 93 PEGs and 51 MEGs. Pairwise comparisons of dN/dS were made between syntenic genes in the genomes of *Zea mays* (v2, id 11266), *Oryza sativa* Japonica (v7, id 16890 masked), and *Sorghum bicolor* (v1.4 id95 masked repeats 50x) using the software SynMap and SynFind. To identify differences in patterns of evolution across codons of conserved imprinted genes, we performed Fast, Unconstrained Bayesian AppRoximation (FUBAR; 46) analyses (see supplemental Methods for additional details).

**Acknowledgments**: Peter Hermanson assisted with the crosses and tissue collections for this study. Ming Luo, Anna Koltunow, and Jen Taylor shared rice imprinting data. The University of Minnesota Biomedical Genomics Center performed the Illumina sequencing for this study. Computational support was provided by the Minnesota Supercomputing Institute and the Texas Advanced Computing Center. This work was created using resources or cyberinfrastructure provided by iPlant Collaborative. The iPlant Collaborative is funded by a grant from the National Science Foundation (#DBI-0735191). URL: www.iplantcollaborative.org. The research was supported by a grant from the National Science Foundation to NMS and MG (MCB-1121952).

Figure Legends

Figure 1. Discovery of imprinted genes in maize. (A) Allele-specific expression analysis for the reciprocal F1 genotypes generated by crossing B73 and Oh43. The proportion of maternal transcripts in both reciprocal hybrids is plotted for the 13,478 genes that had at least 10 allelic reads in both directions of the cross of B73 and Oh43. Similar plots for the other reciprocal hybrids are shown in Fig. S2. Circle symbols represent genes that are significantly ($\chi^2$<0.05) different from the expected 2:1 maternal to paternal ratio whereas square symbols are genes that do not significantly differ from expected ($\chi^2$>0.05). The pink shaded areas indicate moderate MEGs and PEGs, ($\chi^2$<0.05 and at least 80% maternal bias or 60% paternal bias, respectively). The blue shaded areas indicate strong imprinting have significant allelic bias ($\chi^2$<0.01) and exhibit at least 90% parental bias for both MEGs and PEGs. The arrows heads indicate genes with complete imprinting (at least 99% parental bias for MEGs and PEGs and $\chi^2$<0.01)). The green shaded areas indicate genes with potential allelic variation for imprinting (are strongly imprinted in one direction of the cross and biallelic in the reciprocal cross). (B) The proportion of moderate (pink), strong (blue) and complete (gray) imprinting for all non-redundant MEGs and PEGs that were detected in at least one of the five reciprocal crosses.

Figure 2. A subset of imprinted genes show endosperm preferred expression while other imprinted genes are expressed in vegetative tissues. The gene expression patterns for the PEGs (A) and MEGs (B) were obtained from the maize gene expression atlas (Sekhon et al., 2013). The normalized values (per gene) were used for hierarchical clustering (Ward's method) and the heat map indicates relative levels of expression (red – high; black – intermediate; blue – low). The genes with preferential expression in endosperm are indicated to the left of each heat map. The whole seed samples are 2, 4, 6, 10 and 14 DAP (left to right). The endosperm samples are 12, 14, 16, 18, 20, 22 and 24 DAP. The embryo samples are 16, 18 and 22 DAP. The vegetative samples are 18 DAP pericarp, anthers, pre-pollination cob, silks, leaves, stem, immature tassel, immature leaves, immature cob, meiotic tassel, first internode, shoot tip, and three leaf stages.

Figure 3- Conservation of imprinting among maize haplotypes. (A) The proportion of expression from the maternal allele using a heatmap (blue = 0; red = 1; yellow = 0.66; gray = missing data) is shown for all ten genotypes for each of the non-redundant imprinted genes. (B) A similar heatmap is shown for the seventeen genes with allelic variation for imprinting. (C) The expression patterns for one of the allele-specific imprinted MEGs (GRMZM2G384780) is shown.  For this gene, the B73 allele is not silenced when paternally inherited but alleles from the other haplotypes are silenced when inherited from the paternal parent.  For each bar, the upper portion represents the proportion of paternal expression and the lower portion represents the proportion of maternal expression (see gray bars in legend for expectations for MEGs, biallelic and PEGs).  The colors represent the four alleles assessed (see legend), and the values listed inside the bars are the number of maternal (M, bottom) or paternal (P, top) reads. The orange dashed line across the plot represents the expected biallelic ratio of 66% maternal reads.  Black boxes highlight the non-imprinted allele. (D) Allele-specific imprinting pattern for the PEG GRMZM2G106222, which exhibits a failure to silence the Oh43 when it is maternally inherited.  This gene was also validated by a quantitative SNP assay.  SNPs were available to distinguish B73-Mo17 and B73-Oh43 alleles.  The values listed above the bars are the proportion of the maternal allele determined from the quantitative SNP assay.

Figure 4.  Genes with conserved imprinting exhibit evidence for positive selection.  (A) Genes with conserved imprinting exhibit differential evidence of selection.  dn/ds values for genome-wide comparisons of maize (M), rice (R), and sorghum (S). In each comparison, the width of the violin plot (white) represents the genome-wide distributions of dn/ds, red dots represent values for non-conserved imprinted genes in maize, and blue dots represent values for genes with conserved imprinting. Because imprinting data is not available in sorghum, the sorghum ortholog of maize imprinted genes was used in the RS comparison.

Table 1. Discovery of maize imprinted genes

|  | B73 / Mo17 | B73 / Ki11 | Mo17 / Ki11 | B73 / Oh43 | Mo17 / Oh43 | All | NR |
|---|---|---|---|---|---|---|---|
| # genes with >= 10 reads | 11,856 | 10,531 | 5,851 | 13,478 | 9,434 | 2,087 | 18,284 |
| Maternal bias | 81 | 58 | 77 | 180 | 134 | 6 | 394 |
| Moderate MEGs | 75 | 42 | 22 | 118 | 44 | 4 | 198 |
| Strong MEGs | 31 | 28 | 9 | 39 | 25 | 4 | 69 |
| Complete MEGs | 13 | 9 | 3 | 16 | 12 | 3 | 37 |
| Paternal bias | 432 | 563 | 403 | 724 | 487 | 24 | 1,750 |
| Moderate PEGs | 171 | 192 | 74 | 191 | 120 | 18 | 367 |
| Strong PEGs | 56 | 55 | 24 | 76 | 45 | 6 | 108 |
| Complete PEGs | 8 | 17 | 3 | 15 | 5 | 0 | 31 |

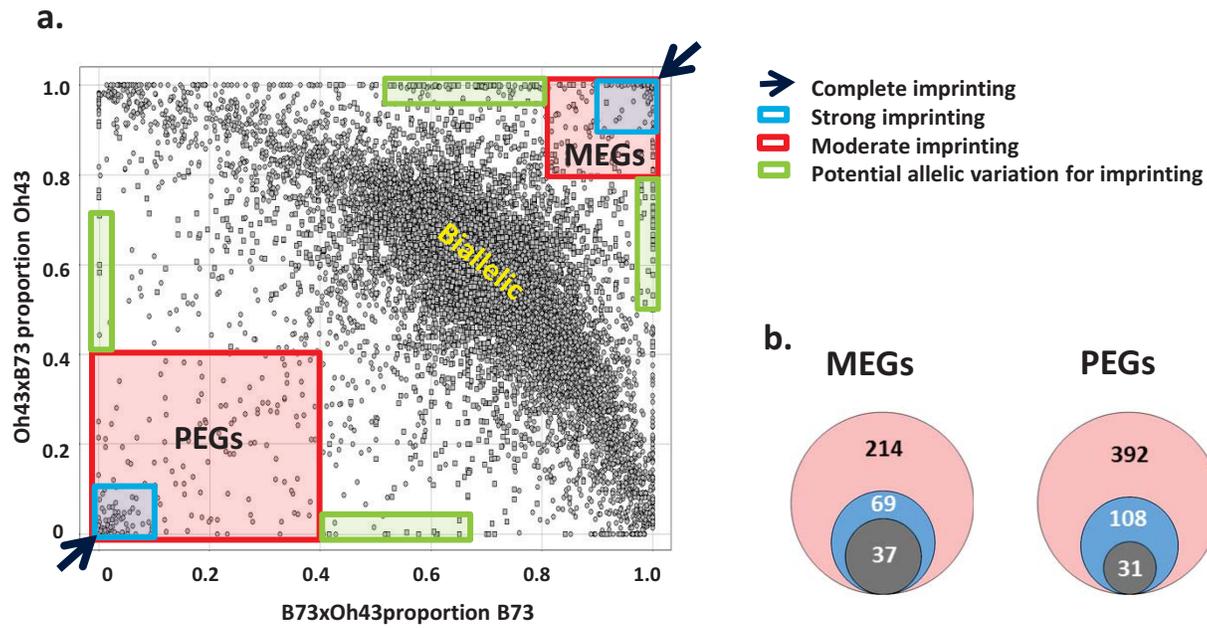

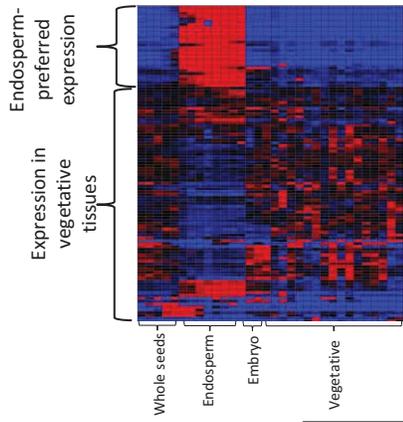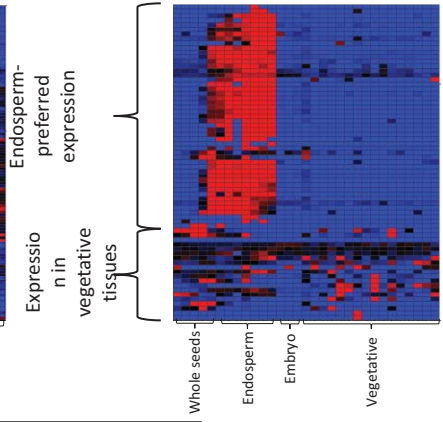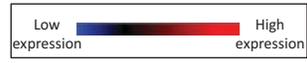

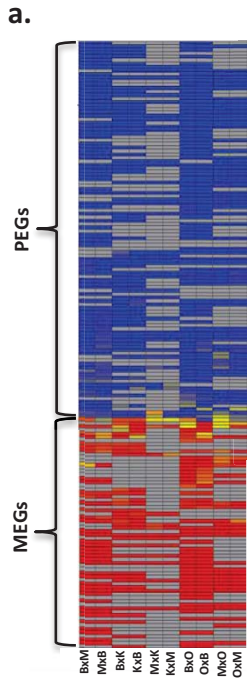
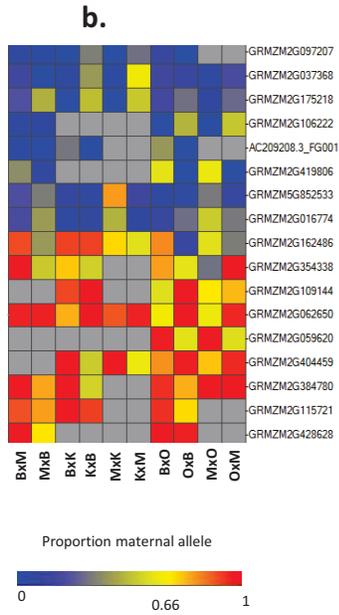
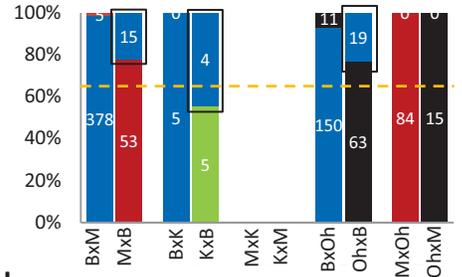
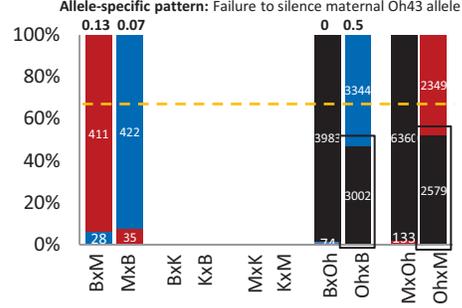
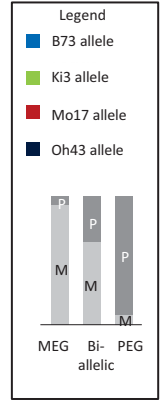

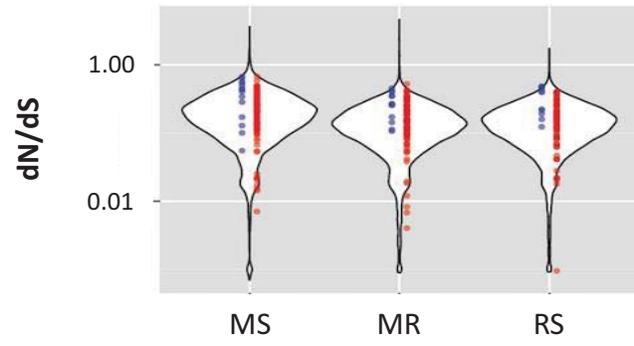

**Supplemental methods**

***Plant materials and RNAseq*** B73, Mo17, Ki11 and Oh43 plants were grown in Saint Paul at the University of Minnesota Agricultural experiment station during the summer of 2011. Reciprocal crosses and self-pollinations for all genotypes were performed between August $4^{th}$ – $15^{th}$ and several ears representing each cross were harvested 14 days after pollination (DAP). The endosperm and embryo tissue were dissected from at least two ears for each genotype and were pooled together and frozen in liquid nitrogen. RNA was isolated by SDS-Trizol protocol and subsequently purified by LiCl precipitations. These RNA samples were submitted to the University of Minnesota Biomedical Genomics Center for sequencing using the Illumina HiSeq-2500 platform.

These reads were aligned to the 39,540 genes in the filtered gene set (version 5b.60) using the Tophat aligner (47) and used to generate relative values for gene expression (FPKM – fragments per kilobase per million reads). The sequence reads were run through an allele specific expression pipeline that aligns reads using Tophat aligner, incorporating SNPs from the HapMap2 project (34) and allowing a maximum of two mismatches per read to assess allele specific expression rates. In order to eliminate potential false-positive SNPs each SNP had to be supported by at least 1% of the reads at that position in the pair of reciprocal hybrids. After filtering there were 28,195 - 142,033 SNPs that were used to assess allele-specific expression in each pair of reciprocal hybrids (Figure S1). The number of reads containing the B73, Mo17, Oh43, or Ki11 allele was summed for all SNPs within the same gene. The alignments were then analyzed to assess the number of reads that align to each of the parental alleles. Genes that have at least 10 allelic reads for each direction of the cross were used to perform chi-square tests (relative to an expected 2 maternal : 1 paternal) ratio. RNAseq reads are deposited at the NCBI SRA under accession (in progress).

***Quantitative SNP assays***: Quantitative SNP assays (Sequenom MassArray) were used to validate imprinted genes and assess imprinting across additional genotypes or over a time course of seed development. cDNA sequences of candidate genes were used to create primers that amplify an informative SNP, which differentiates between two inbred lines. Genomic DNA from parental lines: B73, Mo17, Ki11, Oh43, Ms71, M162W, and NC358; endosperm cDNA from reciprocal crosses : B73xOh43, Mo17xOh43, B73xKi11, Mo17xKi11, B73xMs71, Mo17xMs71, B73xM162W, Mo17xM162W harvested 14 DAP; and endosperm cDNA from reciprocal crosses: B73xMo17, B73xNC358, and Mo17xNC358 harvested 8 DAP, 14 DAP, 16 DAP, 18 DAP, and 20 DAP were submitted to the University of Minnesota BioMedical Genomics Center. The imprinting status was then assessed across all genotypes and time points using standard Sequenom assay conditions (49).

***Population genetic analyses***: Comparisons of diversity statistics between groups of genes were made in R (50) using a Wilcoxon rank sum test. Genome wide substitution rates were calculated using the software on the CoGe website (http://genomevolution.org/). Only the top syntenic hit for each comparison was used in subsequent analyses and dN/dS values above 6 were discarded as likely spurious hits. We further filtered the data to only include those genes for which differences in expression were detectable in our maize dataset (see above). Only those genes and their syntenic orthologs (syntelogs) were kept. To identify the substitution rate for *O. sativa* - *S. bicolor* comparisons, we kept only the genes that shared a common ortholog in maize. Duplicate entries of rice and sorghum genes due to multiple maize orthologs were random pruned.

To identify differences in patterns of evolution across codons of conserved imprinted genes, we performed Fast, Unconstrained Bayesian AppRoximation (FUBAR; 48) analyses. Of the 15 genes with conserved imprinting (12 with rice syntelogs and 3 with non-syntenic homologous rice genes), syntelogs were identified using SynFind software in maize, rice, sorghum, *Brachypodium distachyon* line Bd21 (v1 id8120), and *Setaria italica* (CNS PL3.0 v2.1 id19491) for all except AC191534.3_FG003 and GRMZM2G108309, for which fewer than 3 syntelogs were identified. Maize paralogs identified by SynFind were retained in the analysis. Protein alignments of these 12 gene families were performed with translation align in Geneious (v.5.4.4) under default settings. Neighbor joining trees for FUBAR analyses were constructed from the protein alignments of each gene family using the online HyPhy analysis package (51) and its default settings as implemented on the server [http://www.datamonkey.org/](http://www.datamonkey.org/). FUBAR analyses were also performed on [http://www.datamonkey.org/](http://www.datamonkey.org/), and were run with default settings. FUBAR was run for each alignment, and a posterior probability cutoff of 0.90 was used to identify sites under positive and negative selection. For comparison, analyses were repeated for a random set of 12 imprinted genes (Figure S6A) lacking conserved imprinting in rice.

Dataset S1: Summary of allele specific read counts and features of strong and complete PEGs.

Dataset S2: Summary of allele specific read counts and features of strong and complete MEGs.

Dataset S3: Summary of allele specific read counts and features of genes that exhibit allelic variation of imprinting.

Supplemental table 1: Validation of imprinted genes using the quantitative SNP assay Sequenom.

Supplemental Table 2: Median Values of nucleotide diversity ($\pi$), haplotype heterozygosity (He), Tajima's D, and normalized Fay and Wu's H (H') for MEGS, PEGS, and the full set of genes evaluated. Values are highlighted in bold if they are significantly different from the full gene set (Wilcoxon rank test p-value<0.05).

Supplemental Table 3: Number of genes found within the maize subgenomes as annotated by Schnable et al 2012. Genes are tested for presence in both subgenomes as well as having reported dN/dS values between the subgenomes.

Supplemental Table 1. Sequenom Validation of Imprinted Genes

A. Sequenom validation of MEGs

| Gene ID | Imprinting Level | B73x Mo17 | B73x Oh43 | Mo17x Oh43 | B73x Ki11 | Mo17x Ki11 | Validation in original genotypes? | B73x NC358 | Mo17x NC358 | B73x M162W | Mo17x M162W | B73x Ms71 | Mo17x Ms71 | Validation in additional genotypes? | Validation in time-course? |
|---|---|---|---|---|---|---|---|---|---|---|---|---|---|---|---|
| GRMZM2G009465 | cMEG | MEG | nd | MEG | nd | nd | Yes | nd | MEG | nd | MEG | nd | MEG | Yes | Yes |
| GRMZM2G014119 | cMEG | MEG | nd | nd | nd | MEG | Yes | MEG | nd | nd | nd | nd | MEG | Yes | Yes |
| GRMZM2G063498 | sMEG | MEG | MEG | nd | nd | nd | Yes | MEG | nd | nd | nd | nd | nd | Yes | Yes |
| GRMZM2G073700 | sMEG | MEG | nd | nd | MEG | nd | Yes | nd | MEG | nd | nd | MEG | nd | Yes | Yes |
| GRMZM2G150134 | cMEG | MEG | MEG | nd | MEG | nd | Yes | MEG | nd | nd | MEG | MEG | nd | Yes | Yes |
| GRMZM2G160687 | cMEG[1] | MEG | MEG | nd | nd | nd | Yes | MEG | nd | ASI | nd | MEG | nd | ASI | Yes |
| GRMZM2G169695 | cMEG | MEG | nd | nd | nd | nd | Yes | nd | MEG | nd | nd | nd | MEG | Yes | Yes |
| GRMZM2G178176 | MEG[2] | MEG | nd | nd | nd | nd | Yes | MEG | nd | nd | nd | MEG | nd | Yes | Yes |
| GRMZM2G345700 | MEG[2] | MEG | nd | nd | nd | nd | Yes | nd | nd | nd | nd | nd | nd | Yes | Yes** |
| GRMZM2G354579 | cMEG | MEG | nd | nd | MEG | nd | Yes | MEG | nd | nd | nd | no | MEG | Yes | Yes |
| GRMZM2G370991 | cMEG | MEG | MEG | nd | nd | nd | Yes | nd | MEG | MEG | nd | MEG | nd | Yes | Yes |
| GRMZM2G374088 | mMEG | MEG | nd | MEG* | nd | MEG | Yes | nd | MEG | nd | nd | nd | ASI | ASI | Yes |
| GRMZM5G802403 | cMEG | MEG | nd | nd | nd | nd | Yes | nd | MEG | nd | nd | nd | nd | Yes | Yes |

B. Sequenom validation of PEGs

| Gene ID | Imprinting Level | B73x Mo17 | B73x Oh43 | Mo17x Oh43 | B73x Ki11 | Mo17x Ki11 | Validation in original genotypes? | B73x NC358 | Mo17x NC358 | B73x M162W | Mo17x M162W | B73x Ms71 | Mo17x Ms71 | Validation in additional genotypes? | Validation in time-course? |
|---|---|---|---|---|---|---|---|---|---|---|---|---|---|---|---|
| GRMZM2G000404 | cPEG | PEG | nd | PEG | PEG | nd | Yes | PEG | nd | nd | PEG | nd | PEG | Yes | Yes |
| GRMZM2G002100 | sPEG | PEG | nd | PEG | nd | PEG | Yes | nd | nd | nd | PEG | nd | PEG | Yes | Yes** |
| GRMZM2G006732 | sPEG | PEG | PEG | nd | PEG | nd | Yes | PEG | nd | nd | PEG | PEG | nd | Yes | Yes |
| GRMZM2G020302 | sPEG | PEG | nd | PEG | nd | PEG | Yes | nd | ASI | nd | PEG | nd | PEG | Yes*** | No |
| GRMZM2G040954 | sPEG | PEG | nd | nd | nd | PEG | Yes | nd | PEG | nd | PEG | nd | nd | Yes | Yes |
| GRMZM2G047104 | cPEG | PEG | nd | PEG | nd | PEG | Yes | nd | PEG | nd | PEG | PEG | nd | Yes | Yes |
| GRMZM2G093947 | cPEG | PEG | PEG | nd | PEG | nd | Yes | nd | PEG | PEG | nd | nd | PEG | Yes | Yes |
| GRMZM2G149903 | cPEG | PEG | PEG | nd | PEG | nd | Yes | PEG | nd | PEG | nd | PEG | nd | Yes | Yes |
| GRMZM2G164314 | sPEG | PEG | PEG | nd | PEG | nd | Yes | nd | PEG | nd | PEG | PEG | nd | Yes | Yes |
| GRMZM2G171410 | mPEG | PEG | nd | PEG | nd | PEG | Yes | nd | PEG | PEG | nd | nd | PEG | Yes | Yes |
| GRMZM2G365731 | cPEG | PEG | nd | PEG | nd | PEG | Yes | nd | nd | nd | PEG | PEG | nd | Yes | Yes** |
| GRMZM2G369203 | cPEG | PEG | nd | nd | PEG | nd | Yes | nd | PEG | nd | nd | nd | PEG | Yes | Yes |
| GRMZM2G440949 | sPEG | PEG | PEG | nd | PEG | nd | Yes | PEG | nd | PEG | nd | PEG | nd | Yes | Yes |

No data (nd) and allele specific imprinting (ASI)
[1] Complete MEG that was filtered out as potential contaminant by Sekhon et al. (2013) expression data, but validated in sequenom
[2] Did not meet the minium read requirment in one direction of reciprocal crosses , but validated in sequenom
* Gene that was potentially ASI from the RNAseq data, but validated as imprinted in sequenom
** Validated across the timecourse for B73xMo17, but had no data in B73xNC358 or Mo17xNC358
*** Exhibited allele specific pattern in one additional genotype

Supplemental Table 2- Population genetics statstics for imprinted genes

*Hprime*

|  | Mean | Median | Wilcoxon | Wilcoxon Excluded |
|---|---|---|---|---|
| Full | -0.1792 | 0.1517 | - | - |
| Excluded[1] | -0.1797 | 0.1517 | - | - |
| PEGS | -0.4701 | -0.1805 | 0.09561 | 0.09584 |
| MEGS | 0.6433 | 1.108 | **0.002783** | **0.002754** |

*TajD*

|  | Mean | Median | Wilcoxon | Wilcoxon Excluded |
|---|---|---|---|---|
| Full | 0.6189 | 0.6432 | - | - |
| Excluded[1] | 0.6184 | 0.6432 | - | - |
| PEGS | 0.5086 | 0.5754 | 0.6222 | 0.6248 |
| MEGS | 0.9746 | 0.8335 | 0.1449 | 0.1438 |

*Hapdiv*

|  | Mean | Median | Wilcoxon | Wilcoxon Excluded |
|---|---|---|---|---|
| Full | 0.7883 | 0.9124 | - | - |
| Excluded[1] | 0.7881 | 0.9124 | - | - |
| PEGS | 0.8341 | 0.9306 | 0.2283 | 0.2278 |
| MEGS | 0.7441 | 0.8828 | 0.1299 | 0.1301 |

*ThetaPi*

|  | Mean | Median | Wilcoxon | Wilcoxon Excluded |
|---|---|---|---|---|
| Full | 0.00716 | 0.00848 | - | - |
| Excluded[1] |  |  | - | - |
| PEGS | 0.007871 | 0.006865 | 0.6026 | 0.6003 |
| MEGS | 0.008799 | 0.007554 | 0.8555 | 0.8565 |

*Dn/Ds*

|  | Mean | Median | Wilcoxon | Wilcoxon Excluded |
|---|---|---|---|---|
| Full | 0.2115 | 0.1785 | - | - |
| Excluded[1] | 0.2111 | 0.1781 | - | - |
| PEGS | 0.2558 | 0.2308 | **0.0006869** | **0.0006137** |
| MEGS | 0.2232 | 0.2763 | **0.0289** | **0.02774** |

[1] Values for MEGs and PEGs were removed and then statistics were recalculated

Supplemental Table 3- Presence of imprinted genes and retained paralogs in subgenomes of maize

|          | Hprime[1] | Tajima D[1] | Hapdiv[1] | ThetaPi[1] |
|----------|-----------|-------------|-----------|------------|
| Genomic  | 0.1517    | 0.6432      | 0.9124    | 0.00176    |
| PEG      | -0.1805   | 0.5267      | 0.9317    | 0.006865   |
| MEG      | **1.108*** | 0.9746     | 0.8828    | 0.007554   |

|             | Presence | | | Retained Paralog | | |
|-------------|----------|-----|-----|------------------|-----|-----|
|             | Waters[2] | MEG | PEG | Waters[2] | MEG | PEG |
| Subgenome1  | 7086  | 15 | 46  | 2307  | 4  | 9  |
| Subgenome2  | 4453  | 16 | 29  | 1997  | 3  | 9  |
| Total       | 11539 | 31 | 75  | 4304  | 7  | 18 |
| Not in      | 5926  | 38 | 37  | 7235  | 24 | 57 |
| Original    | 17465 | 69 | 112 | 11539 | 31 | 75 |

|         | Dom | Nodom | masked |       |       |
|---------|-----|-------|--------|-------|-------|
| Waters  | 677 | 13745 | 344    | 14766 | 14089 |
| PEG     | 4   | 88    | 1      | 93    | 89    |
| MEG     | 2   | 46    | 1      | 49    | 47    |

Mattlist

|           | Dom | Improv | Total[3] | Not Dom | Not Improv |
|-----------|-----|--------|----------|---------|------------|
| Waters    | 192 | 93     | 14982    | 14790   | 14889      |
| PEG       | 2   | 1      | 93       | 91      | 92         |
| MEG       | 0   | 0      | 51       | 51      | 51         |
| Imprimted | 2   | 1      | 144      | 142     | 143        |

[1] Values are medians
[2] # genes with >10 allele specific reads in both directions of at least one set of reciprocal crosses
[3] Total number of genes in Refgen 1
*Stastically different from genomic median (Wilcoxon rank sum test p-value=0.0028)

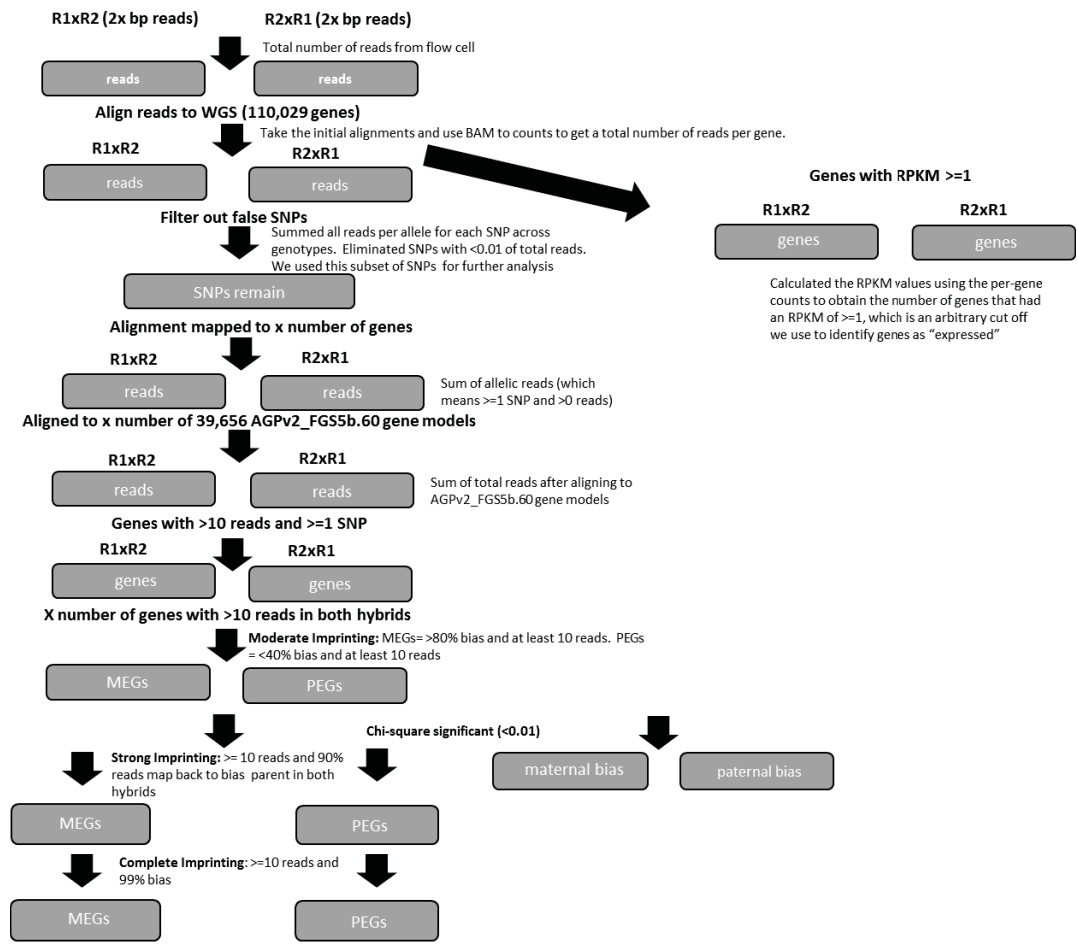

| | B73xMo17 | Mo17xB73 | B73xKi11 | Ki11xB73 | Mo17xKi11 | Ki11xMo17 | B73xOh43 | Oh43xB73 | Mo17xOh43 | Oh43xMo17 |
|---|---|---|---|---|---|---|---|---|---|---|
| Number of raw RNA-seq reads | 209,874,775 | 201,851,318 | 198,126,556 | 192,953,028 | 197,192,383 | 198,328,755 | 197,126,727 | 190,605,935 | 203,700,260 | 179,299,550 |
| Per gene read counts | 60,432,048 | 61,450,235 | 48,761,575 | 138,160,523 | 87,590,543 | 105,458,847 | 206,600,366 | 247,736,952 | 145,469,649 | 161,979,127 |
| RPKM >1 | 10,818 | 11,031 | 9,545 | 9,682 | 5,931 | 5,690 | 10,148 | 10,136 | 8,036 | 8,713 |
| # SNPs used | 115,813 | | 69,891 | | 28,195 | | 142,033 | | 73,895 | |
| # Genes with SNPs | 17,262 | | 15,355 | | 8,329 | | 19,504 | | 13,823 | |
| # Reads aligned to genes | 17,134,491 | 15,681,619 | 15,620,862 | 22,676,835 | 14,936,746 | 13,409,826 | 99,084,549 | 84,317,265 | 40,865,059 | 35,523,692 |
| #Genes in FGS | 14,405 | | 15,355 | | 7,092 | | 15,648 | | 11,443 | |
| # Reads aligned to FGS | 14,427,498 | 13,552,279 | 12,930,459 | 19,086,835 | 12,830,641 | 11,160,346 | 76,529,348 | 67,101,931 | 35,638,015 | 27,879,407 |
| Number of genes with >= 10 reads | 12,201 | 12,381 | 10,693 | 11,326 | 6,298 | 6,044 | 13,988 | 13,873 | 10,344 | 9,641 |

Supplemental Figure 1- Analysis pipeline for RNA-seq based discovery of imprinted genes.  Raw reads from reciprocal hybrids were aligned to the working gene set of version two of the B73 reference genome  using TopHat aligner.  Two different alignment iterations were completed; allele specific expression analysis (the flowchart) and gene specific alignments (calculate RPKM values).  SNPs from resequence data were used to asses allele specific expression rates (Chia et al, 2011). We required at least 1% of the allelic reads for any given SNP came from both parents.  False SNPs have all reads map to one parental allele in both reciprocal crosses.  Once false SNPs  were filtered out the allelic reads covering the remaining SNPs (variant reads) were summed for each gene.  Genes in the filtered gene set (AGPv2_FGSv5b) were used for further analysis.  At least 10 variant reads are required in both directions of the cross to assess parental bias expression.   A chi-square significance test was preformed on each gene.  Maternally and paternally biased genes  have a significant parental bias ( <0.01).  Additional cutoffs  were used to identify and categorize MEGs and PEGs into three categories: moderate, strong, and complete.   Moderate MEGs and PEGs have a significant parental bias expression ( <0.05) and >80% maternal bias or >60% paternal bias, respectively.  Strong MEGs and PEGs have a significant parental allelic bias ( <0.01) and >90% maternal or paternal bias, respectively.  Complete MEGs and PEGs have >99% maternal or paternal bias, respectively.

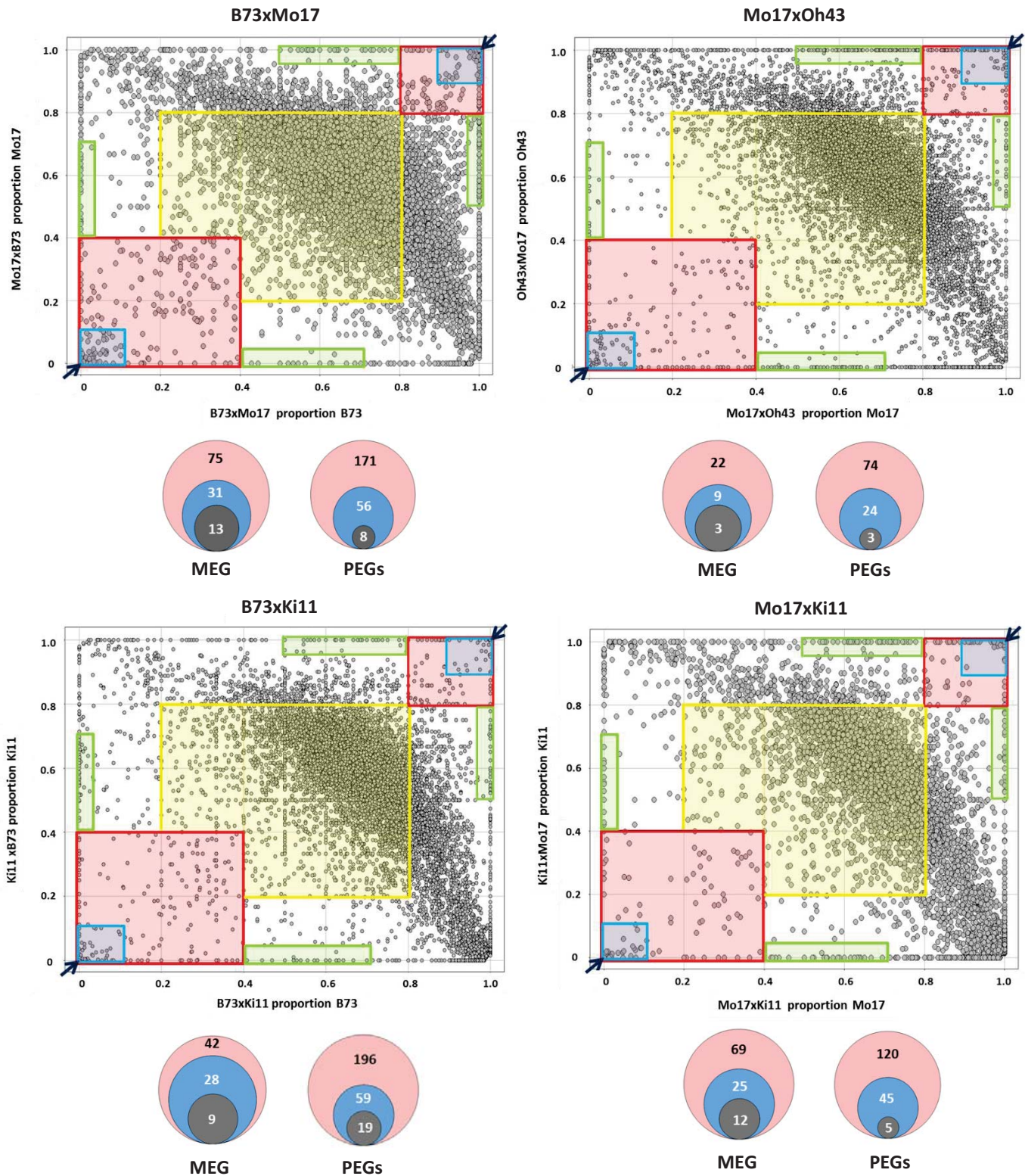

Supplementary Figure 2- Proportion of maternal transcripts and number of moderate, strong and complete MEGs and PEGs for B73xMo17, Mo17xOh43, B73xKi11, and Mo17xKi11. Moderate MEGs and PEGs (>80% maternal or >60% paternal bias, respectively) are within the red boxes. Strong MEGs and PEGs (>90% parental bias) are within the blue boxes. Complete MEGs and PEGs (>99% parental bias) are represented by the arrows. Biallelic genes are within the yellow squares. The green boxes represent allele specific imprinted genes The number of moderate, strong, and complete MEGs and PEGs for each genotype are represented by the red, blue, and grey circles, respectively.

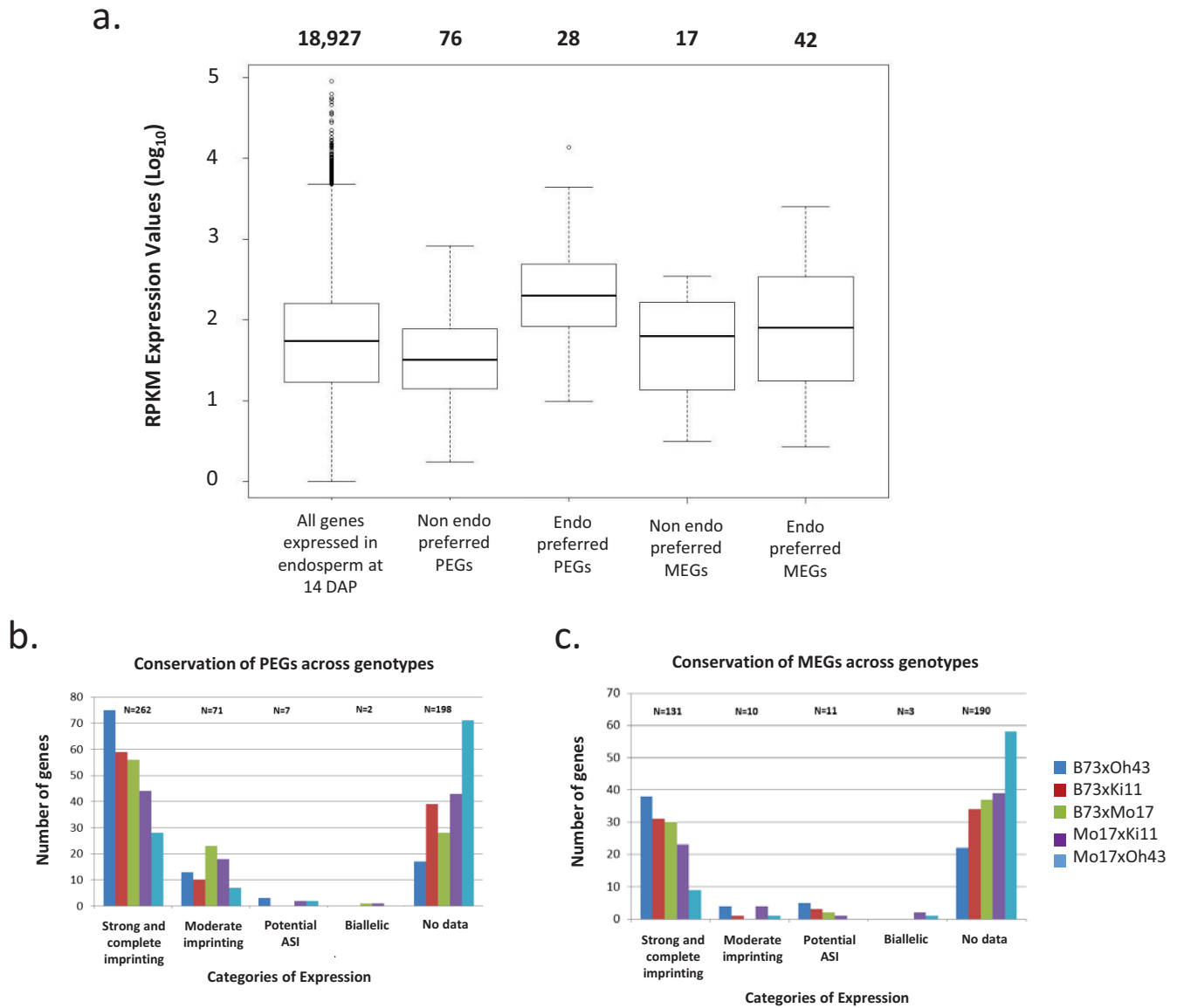

Supplementary Figure 3-Comparison of tissue specific expression and conservation of imprinting among maize haplotypes. (A) RPKM expression values are log transformed data from Sekhon et al., 2013. All genes with RPKM values of 0 were removed from this analysis. Gene expression was compared across all genes that are expressed in endosperm tissue at 14 DAP to MEGs and PEGs that are either preferentially expressed in endosperm or show equivalent expression in other tissue types. The values listed above box plots are the number of genes in each group. PEGs that are endosperm preferred have a higher average expression 2.3) relative to all other groups (1.5-1.8). No statistical difference in expression was observed for either group of MEGs or PEGs that are not preferentially expressed in endosperm. (B) The conservation level of imprinting was assessed within maize genotypes. Very few genes identified as MEGs and PEGs in one pair of crosses is biallelic in another set of crosses (0.009 and 0.004, respectively). Each color represents a pair of reciprocal F1 hybrid crosses. Each non-redundant imprinted gene could be placed in one of five categories for each pair of reciprocal crosses: Strong imprinting (>90% parental bias), moderate imprinting (>80% maternal bias (MEGs) or >60% paternal bias (PEGs)), potentially allele specific imprinting (ASI, strong imprinting in one direction and biallelic expression in the other direction), or no data (no SNPs or <10 reads). The values above the chart show the number of genes that fit in each category. Over 97% of PEGs and 91% of MEGs are at least moderately imprinted in all genotypes with data. A lack of polymorphisms or reads (no data call) is more prevalent for MEGs(0.55) than PEGs (0.37).

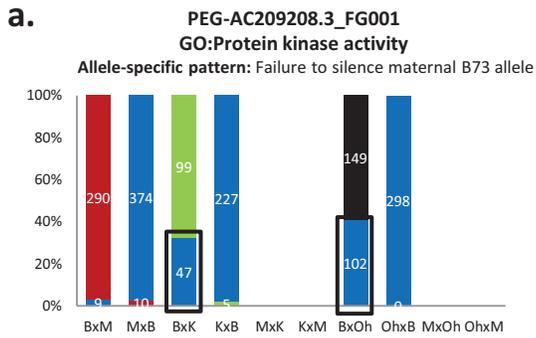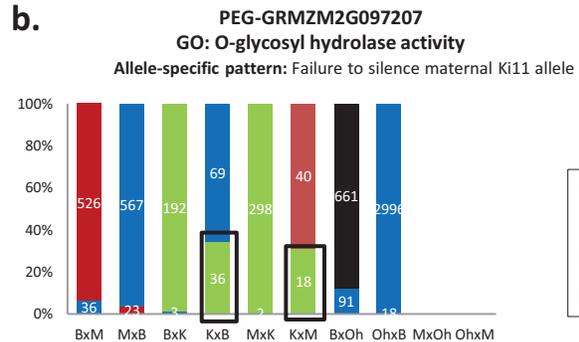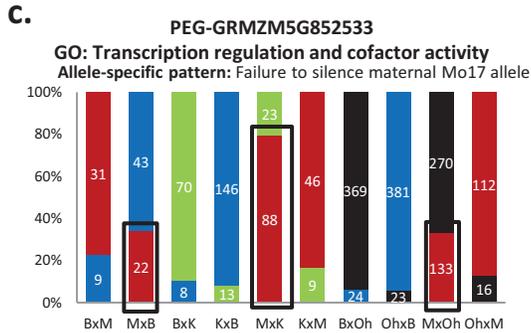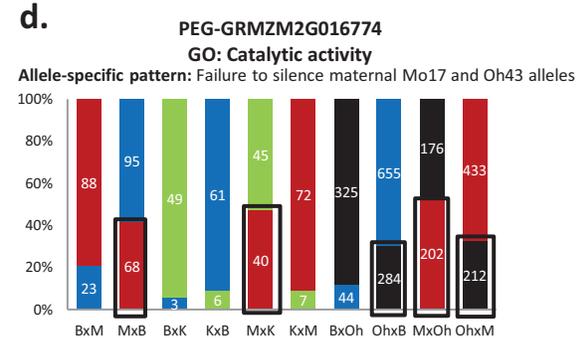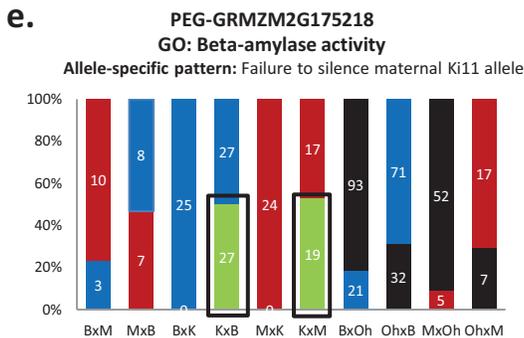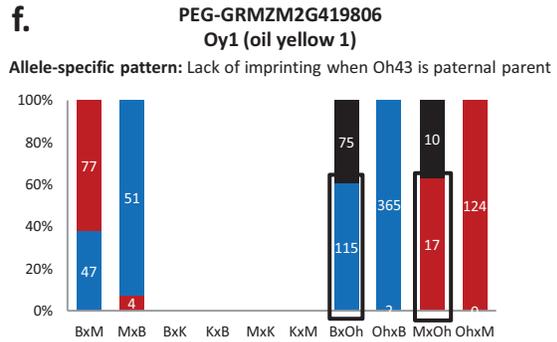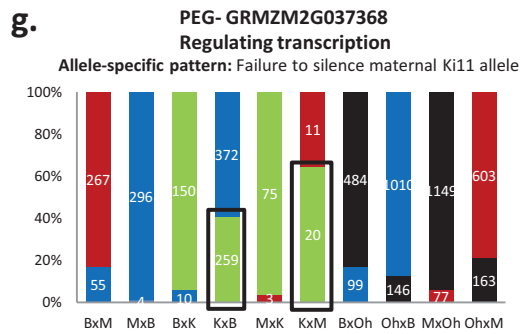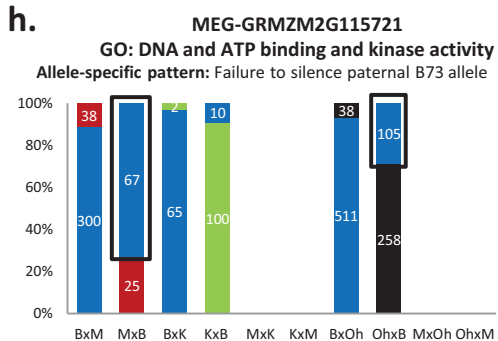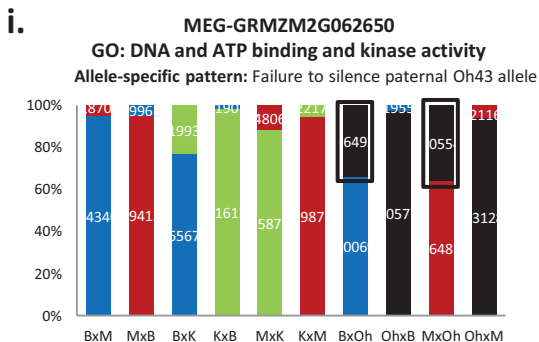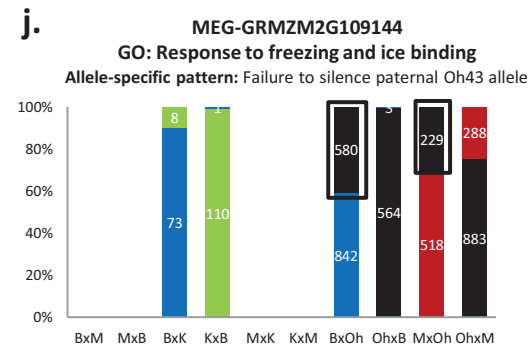

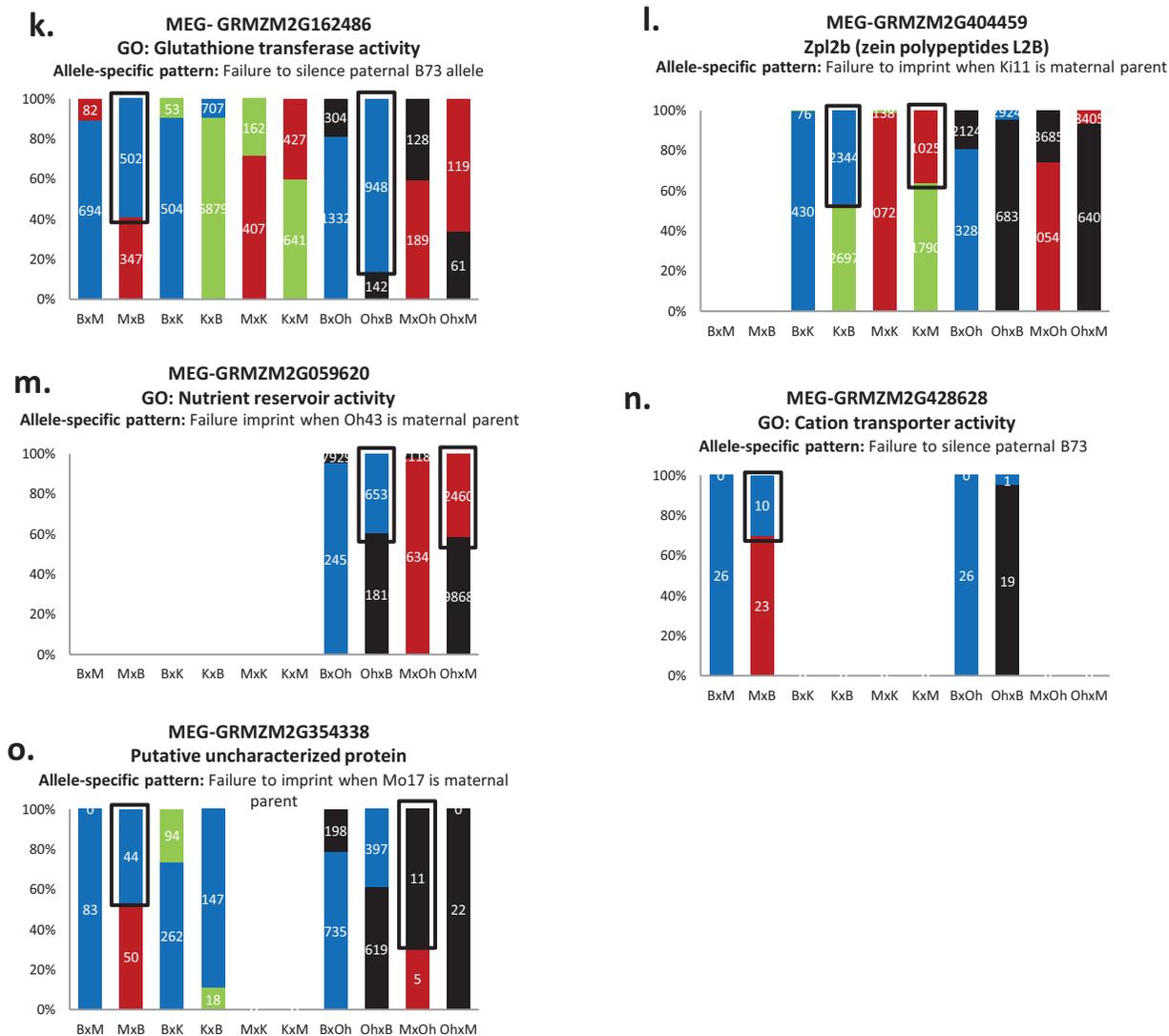

Supplementary Figure 5: PEGs and MEGs exhibit allelic variation of imprinting. The number of maternal reads (top) and paternal reads (bottom) are shown for each of the seven PEGs and eight MEGs that exhibit allelic variation of imprinting. Each color represents a different allele and the black boxes highlight the allele(s) that fails to silence. A variety of patterns were observed in terms of number of alleles and which alleles fail to silence. Many of these PEGs fail to silence certain alleles when they are maternally inherited (**A-E** and **G**), only 1 example (**F**) the allele fails to silence when it is paternally inherited. In addition, one allele fails to silence in a majority of these PEGs (**A-C, E and G**), whereas in **D and F** multiple alleles fail to silence when maternally or paternally inherited, respectively. B73 fails to silence in **A** and **F**, Mo17 fails to silence in **C**, **D**, and **F**, Ki11 fails to silence in **B**, **E** and **G,** and Oh43 fails to silence in **F**. Characterized gene oil yellow exhibits allelic variation of imprinting, in that both B73 and Mo17 fail to silence when Oh43 is inherited paternally. Similar patterns were observed for MEGs that exhibit variation of imprinting. The B73 allele failed to silence when it paternally inherited in **H, K,** and **N** and when maternally inherited in **M** and **O**. The Oh43 allele failed to silence when paternally inherited in **I** and **J**, and when maternally inherited in **O**. Failure to silence Mo17 was only observed when Mo17 was inherited maternally and additional alleles failed to silence when Ki11 and Oh43 are maternally inherited (**L** and **M**, respectively).

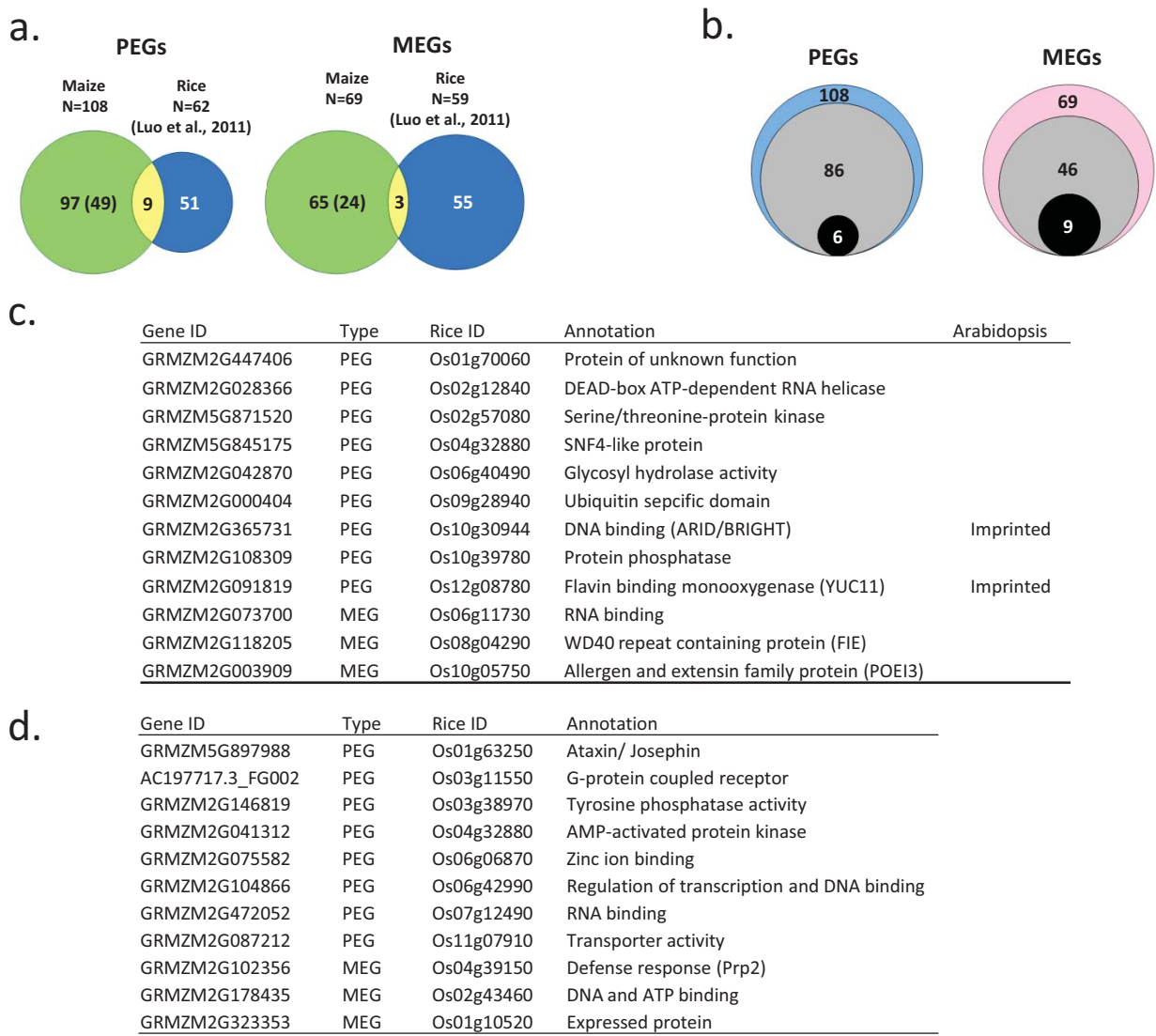

Supplemental Figure 5: Limited, but significant, conservation of imprinting between maize, rice and *Arabidopsis*. Syntenic regions in rice were identified and the level of conservation was assessed (A). The values in parentheses are the number of maize imprinted genes that have a syntenic ortholog in rice and were assessed by Luo et al (2011). Assessing conservation of syntenic genes identified 9 PEGs and 3 MEGs that are conservatively imprinted between maize and rice. (B) The conservation of imprinting between maize and rice was assessed. There are an additional 2 PEGs and 1 MEG that have highly related rice homologs that are imprinted but not located in syntenic genomic positions. (B) The level of conservation between maize and *Arabidopsis thaliana* was assessed. The protein sequence of each maize imprinted gene was used to find potential orthologs in *Arabidopsis* (BLASTp). The blue outer circle is the number of non-redundant maize PEGs and the pink outer circle is the number of non-redundant maize MEGs. The grey circles are the number of PEGs (0.80) or MEGs (0.67) that have a potential ortholog in *Arabidopsis* (BLAST e-value <1e-20) for which the Arabidopsis ortholog was assessed by either Gehring et al (2011) or Heish et al (2011). The black circles are the number of MEGs or PEGs that are conservatively imprinted between maize and *Arabidopsis*. Approximately, 7% of PEGs compared to 20% of MEGs, for which an ortholog was assessed are conserved between maize and *Arabidopsis*. (C) The maize gene ID, rice gene ID, annotation and imprinting status in *Arabidopsis* for PEGs and MEGs that are conservatively imprinted in maize and syntenic blocks in rice. (D) Maize gene ID, rice gene ID, and annotation for the 3 moderate MEGs and 8 moderate PEGs that show conservation of imprinting between maize and syntenic blocks in rice.

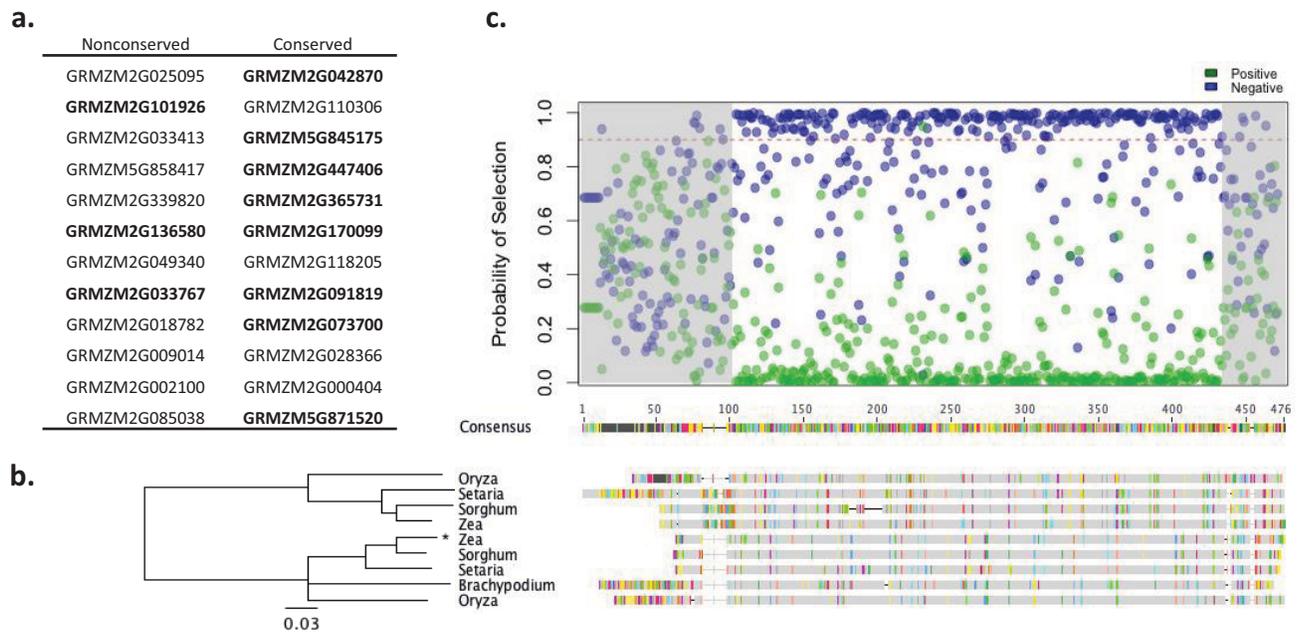

Supplementary Figure 8: Genes with conserved imprinting exhibit evidence for positive selection. (A) Gene IDs for which FUBAR analyses were performed. Gene IDs in bold showed at least one codon under positive selection. Nonconserved imprinted genes represent a single random sample.
(B) Alignment and neighbor-joining tree of the imprinted maize gene GRMZM2G042870 (shown with asterisk) and syntenic orthologs in the grasses. Colored tick marks in the alignment indicate mismatches to the consensus sequence. (C) MEME analysis (see methods) of posterior probability of positive (green) and negative (blue) selection along the coding sequence of GRMZM2G042870. The dashed line indicates 90% posterior probability. Regions of low confidence in the alignment in indicated with a transparent gray box.